\documentclass[aps,prb,onecolumn,11pt]{revtex4-2}
\usepackage{amsmath,amssymb}
\usepackage[T1]{fontenc}
\usepackage{lmodern}
\usepackage{hyperref}
\begin{document}
\title{Comment on ``Is there a density-gradient expansion for the electronic exchange energy?''}
\author{Mario Benites}
\author{\'{A}ngel Rosado}
\author{Efstratios Manousakis}
\affiliation{Department of Physics, Florida State University, Tallahassee, Florida 32306, USA}
\begin{abstract}
  Perdew et al.~\cite{Perdew} dispute our finding~\cite{Benites} that the separate exchange ($b_x$) and correlation ($b_c$) contributions to the $q^2$ coefficient of the long-wavelength proper polarization function depend on the infrared regularization of the Coulomb interaction, although their sum $b_{xc}=b_x+b_c$ does not. We question their claim that the conventional exchange coefficient is uniquely determined. In Qian's~\cite{Qian} derivation, which Perdew et al.~\cite{Perdew} cite in support of their claim of a unique exchange-gradient coefficient, an integral of the form $\int_{-a}^{b}dx/x$ crosses a singularity and does not exist as an ordinary improper integral. Its finite Cauchy principal value (which is
  a particular way of regularizing the integral) is obtained only by excluding equal intervals on the two sides of the singularity; unequal exclusions give different answers. Independently, Sham's~\cite{Sham} calculation with the screened interaction $e^{-\lambda_0 r}/r$ gives an exchange-gradient coefficient different from the bare-Coulomb calculations of Svendsen--von Barth~\cite{Svendsen} and Engel--Vosko~\cite{Engel}. The discrepancy remains when $\lambda_0>0$ is so small that $1/\lambda_0$ is comparable to the size of the observable universe, making the two interactions experimentally indistinguishable on laboratory scales. Why should the exactly unscreened result be preferred, even if its calculation is accepted as valid? Our calculation shows that the ambiguity of the separate contributions cancels in the combined, regulator-independent coefficient $b_{xc}$.
\end{abstract}
\maketitle

Perdew et al.~\cite{Perdew} argue that taking the bare-Coulomb limit first gives the uniquely correct exchange-gradient coefficient $C_x$, invoking the calculations of Svendsen and von Barth, Engel and Vosko, and Qian~\cite{Svendsen,Engel,Qian}. Agreement among such results does not establish uniqueness without examining their singular integrals and limiting prescriptions. Moreover, Sham's density-independent screened-Coulomb calculation~\cite{Sham} gives a different value; this discrepancy predates our density-dependent regularization~\cite{Benites}.

Consider Qian's calculation~\cite{Qian} of the lowest-order correction $\Pi_1(q,0)$ to the Lindhard function, which Perdew et al.~\cite{Perdew} cite in support of their claim of a unique exchange-gradient coefficient. Its original six-dimensional momentum integral given by Eq.~(5) of Ref.~\cite{Qian}, contains terms proportional to $(\mathbf p\cdot\mathbf k)^{-2}$. The reduction produces one-dimensional integrals of the form (see the text before Qian's Eq.~(46))
\begin{equation}
\int_{-a}^{b}\frac{dz'}{z'}, \qquad a,b>0.
\label{eq:singular}
\end{equation}
The interval of integration has a crossing point $z'=0$ at the denominator level, so this integral does not exist as an ordinary improper integral~\cite{Apostol,Rudin,Whittaker}. At both sides of the interval, the integral diverges separately. To  obtain a finite result requires an additional prescription, such as the Cauchy principal value,
\begin{equation}
\operatorname{PV}\int_{-a}^{b}\frac{dx}{x}
\equiv\lim_{\epsilon\to0^+}\int_{-a}^{b}dx\,I_\epsilon(x),
\label{eq:pv}
\end{equation}
where
\begin{equation}
I_\epsilon(x)=\begin{cases}
1/x,& |x|\ge\epsilon,\\
0,& |x|<\epsilon.
\end{cases}
\label{eq:cutoff}
\end{equation}
This prescription gives $\ln(b/a)$, and zero when $a=b$. But independently approaching the singularity from the two sides gives
\begin{equation}
\int_{-a}^{b}\frac{dx}{x}\ \overset{\mathrm{prescription}}{=}\
\lim_{\substack{\epsilon_1\to0^+\\\epsilon_2\to0^+}}
\left[\int_{-a}^{-\epsilon_1}\frac{dx}{x}
+\int_{\epsilon_2}^{b}\frac{dx}{x}\right].
\label{eq:asymmetric}
\end{equation}
The bracket equals $\ln(b/a)+\ln(\epsilon_1/\epsilon_2)$. For $a=b$, setting $\epsilon_1=2\epsilon_2$ yields $\ln2$ rather than zero. Thus the value depends on the prescription. The issue arises in Qian's terms, $J_0$ and $J_{23}=J_2+J_3$ given by Eqs.~(12) and (44) in Ref.~\cite{Qian}, which enter $\Pi_1(q,0)$ through Eqs.~(6) and (11). Qian does not justify the treatment of these singular terms as convergent ordinary integrals. What, then, warrants regarding the coefficient extracted from that calculation as uniquely established?

Antoniewicz and Kleinman~\cite{AntoniewiczKleinman} explicitly state, immediately below their Eq.~(9), that they numerically evaluated the \emph{principal value} of their singular integral, specifying a mesh prescription to do so. They also report that their calculation becomes numerically unstable as $\kappa=K/k_F\to0$ and describe $\kappa=0$ as an essential singularity of the ratio used to extract the gradient coefficient, where that ratio is undefined. They therefore obtain the coefficient by extrapolating numerical results toward $\kappa=0$, rather than by evaluating an ordinary integral there. This is further evidence that the conventional exchange-gradient coefficient relies on a prescribed treatment of the singular limit.

The concern is not restricted to Qian's derivation. Perdew et al.~\cite{Perdew} suggest that our regulator dependence in our exchange gradient coefficient $C_x$ results from using a density-dependent regulator. Yet our family of regulators $\lambda(k_F)$ include the constant $\lambda_0$, independent of density, as in Sham's calculation~\cite{Sham}. Consider the screened-Coulomb integral
\begin{equation}
I_1(\lambda_0)=\int d\mathbf r\int d\mathbf y\,\frac{e^{-\lambda_0 r}}{r}f(\mathbf r,\mathbf y),
\label{eq:sham}
\end{equation}
and the corresponding bare-Coulomb integral employed by Svendsen and von Barth,
\begin{equation}
I_2=\int d\mathbf r\int d\mathbf y\,\frac{1}{r}f(\mathbf r,\mathbf y).
\label{eq:bare}
\end{equation}
For any finite range of physical distances, the potentials can be made indistinguishable to arbitrary experimental accuracy by choosing $\lambda_0$ sufficiently small. One may, for example, take a screening length ($1/\lambda_0$) comparable to or much larger than the size of the observable universe; over a laboratory sample $e^{-\lambda_0 r}$ is then effectively unity. Nevertheless, Sham's screened calculation and the bare-Coulomb calculation yield different values for the exchange-gradient coefficients.

Even if we grant that the bare-Coulomb integral $I_2$ exists and has been evaluated correctly, there remains a decisive question. If $\lim_{\lambda_0\to0^+}I_1(\lambda_0)\ne I_2$, why should the result obtained by setting $\lambda_0=0$ before integration be preferred over the result obtained from arbitrarily small positive $\lambda_0$? A finite answer from either procedure does not prove its uniqueness or its physical priority. The discrepancy shows that the separate coefficient is not independent of how the Coulomb limit is defined. The authors must supply a reason for selecting one prescription; an assertion that the bare interaction has $\lambda_0=0$ is not such a reason. In a material, screening offers an additional physical reason not to regard strictly zero screening as privileged.

Ref.~\cite{Perdew} further argues that derivatives of our density-dependent regulator with respect to $k_F$ are absent from the final combined coefficient $b_{xc}$. In fact these terms are retained and cancel when the contributions are combined: Eq.~(63) and Eqs.~(76)--(78) of Ref.~\cite{Benites} give Eq.~(81). In the proper-polarization response, consistent use of any regulator $R$ in the exchange and correlation parts gives
\begin{equation}
 b_x^{(R)}+b_c^{(R)}=b_{xc}.
\label{eq:sum}
\end{equation}
Here $b_{xc}$ is the coefficient of the $q^2$ term in the long-wavelength expansion of the proper polarization function, characterizing the response to a slowly varying external perturbation. It is regulator independent, although the individual exchange and correlation contributions depend on $R$. These separate contributions arise from a perturbative decomposition and need not have independently defined, regulator-independent values. Their sum does: this cancellation is our central physical result.

Our scheme reproduces the exchange-gradient coefficients reported by Sham and subsequent authors~\cite{Sham,Svendsen,Engel,Qian}, as well as the correlation-gradient coefficients obtained by Ma--Brueckner and Geldart--Rasolt~\cite{Ma,Geldart}. We therefore ask Perdew et al.~\cite{Perdew} three specific questions. First, since Qian's calculation, which Perdew et al.~\cite{Perdew} cite in support of their claim of a unique exchange-gradient coefficient, implicates a singular integral with no ordinary improper-integral value, how can its principal-value evaluation establish a unique exchange coefficient rather than one fixed by a particular regularization? Second, take Sham's density-independent screening parameter to be positive but so small that $1/\lambda_0$ is comparable to the size of the observable universe. The screened and bare interactions are then indistinguishable over laboratory distances, yet Sham's exchange-gradient coefficient differs from the bare-Coulomb result quoted by Svendsen--von Barth and Engel--Vosko. Why is this discrepancy not a concern? Third, even if we grant that the integral evaluated directly with $\lambda_0=0$ exists and is calculated correctly, what justifies choosing that result over $\lim_{\lambda_0\to0^+} I_1(\lambda_0)$ when the two differ? A calculation giving a finite answer does not resolve this choice. By contrast, our central result is that the regulator dependence of the separate exchange and correlation contributions cancels in $b_{xc}=b_x^{(R)}+b_c^{(R)}$, the coefficient governing the complete long-wavelength response.
\iffalse

Our density-dependent regularization also reproduces the correlation-gradient coefficient of Geldart and Rasolt~\cite{Geldart}, despite the criticism of density-dependent regulators.
If the exchange coefficient is taken from Svendsen--von Barth, Engel--Vosko, or Qian, why combine it with the Geldart--Rasolt correlation coefficient (who use Sham's regulator as shown in Ref.~\cite{Benites}) without demonstrating that both contributions are evaluated using a consistent regularization, as required for the cancellation in Eq.~(\ref{eq:sum})?
\fi

\end{document}